# Are micro and macro arrows of time connected each other?

**Piero Chiarelli**[1]

National Council of Research of Italy, Area of Pisa, 56124 Pisa, Moruzzi 1, Italy

E-mail: pchiare@ifc.cnr.it.

**Abstract.** The connection between the micro and macro-arrow of time is discussed in the frame of the stochastic quantum hydrodynamic analogy (SQHA). The presence of fluctuations that in the case on non-linear interactions leads to the breaking of the quantum mechanics on large scale and then to the macroscopic irreversibility with a defined arrow of time, gives also rise to the time reversal breaking in the micro-scale quantum evolution (micro-arrow of time). The quantum irreversibility with time reversal asymmetry is briefly discussed.

## 1. Introduction

The breaking of quantum coherence and the quantum to classical phase transition are problems that have many consequences in all problems of physics whose scale is larger than that one of small atoms.

Both the equations of quantum and the classical mechanics own the time-reversal symmetry. The arrow of time in the classical kinetics is commonly ascribed to the transition from the molecular scale to the macroscopic one united to the presence of a small initial incertitude (or to the external environment) in a system with a huge number of non-linearly interacting particles [1]. Even if the time reversal symmetry breaking has been observed for the probability transition in some elementary particle decay [2], the macroscopic arrow of time has no connection with the nature of the quantum mechanics. Recently the author has developed the stochastic version of the quantum hydrodynamic analogy (SQHA) [3] that comprehends both the standard quantum mechanics as the deterministic limit of the theory and the classical stochastic mechanics as the large-scale limit. Close to the deterministic limit the unified SQHA dynamics can lead to a marginal time reversal breaking (micro-arrow of time) as well as to irreversible kinetics in macroscopic systems of non-linearly interacting particles.

## 2. The SQHA equation of motion

When the noise is a stochastic function of the space, the particles density (PD) $n$ in the SQHA is described by the motion equation [3]

$$\partial_t n_{(q,t)} = -\nabla_q \bullet ( n_{(q,t)} \dot{q} ) + y_{(q,t,\Theta)} \qquad (1)$$

where $\Theta$ is the amplitude of the spatially distributed noise $y$ whose variance reads

$$\lim_{\Theta \to 0} <y_{(q_r,t)}, y_{(q_s+\lambda,t+\tau)}> = k\Theta \frac{\tilde{\ }}{2\lambda_c^2} exp[-(\frac{\lambda}{\lambda_c})^2]\delta(\tau)\delta_{rs} , \qquad (2)$$

where the noise correlation distance $\lambda_c$ reads

$$\lim_{\Theta \to 0} \}_c = f \frac{\hbar}{(2mk\Theta)^{1/2}}, \tag{3}$$

and where $\dot{q}$ in eq.(1) is obtained by the solution of the differential equations

$$\dot{q} = \frac{p}{m}, \tag{4}$$

$$\dot{p} = -\nabla(V_{(q)} + V_{qu(n)}), \tag{5}$$

where $V_{(q)}$ represents the Hamiltonian potential, $V_{qu(n)}$ is the so-called (non-local) quantum potential [3] that reads

$$V_{qu} = -\left(\frac{\hbar^2}{2m}\right) n^{-1/2} \nabla \cdot \nabla n^{1/2} \tag{6}$$

and $\sim$ is the PD mobility that depends by the specificity of the considered system [3]. The noise amplitude parameter $\Theta$ equals the temperature of an ideal gas at equilibrium with the vacuum fluctuations [3]. It is worth noting that equation (1) is the hydrodynamic-like representation of the Schrödinger stochastic equation (SSE) [4]

$$i\hbar \frac{\partial Œ}{\partial t} = -\frac{\hbar^2}{2m} \nabla_q^2 Œ + V_{(q)} Œ + i \frac{Œ}{|Œ|^2} Y_{(q_r, t, \Theta)}. \tag{7}$$

The noise variance (2) is a direct consequence of the quantum potential term $\nabla \cdot \nabla n^{1/2}$ that gives to the system energy a membrane elastic-like contribution [5] tha reads

$$\overline{H}_{qu} = \int_{-\infty}^{\infty} n_{(q,t)} V_{qu(q,t)} dq, \tag{8}$$

where higher curvature of the PD leads to higher energy (see figure 1). Therefore, independent fluctuations infinitesimally far apart leading to vanishing curvature wrinkles of the PD (and hence to an infinite quantum potential energy) are not allowed.. If we require that any physical system can attain configurations with finite energy, independent fluctuations on shorter and shorter distance have smaller and smaller amplitude, leading to the existence of a correlation distance (let's name it $\}_c$) for the noise.

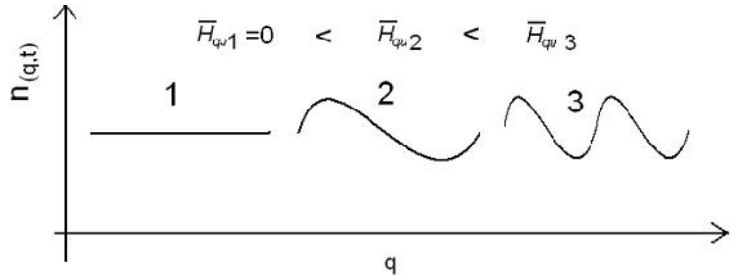

Fig. 1 Quantum potential energy as a function of the "curvature" of the particle density n.

*2.1. Range of interaction of quantum pseudo-potential*

For Hamiltonian potential weaker that the quadratic one (ε<0 in fig. 2), the quantum potential force (QPF) at large distance grows less than linearly [2]. Given the case $lim_{q\to\infty} V_{(q)} \propto q^k$, for $k < \frac{2}{3}$ the QPF at large distance is vanishiong and the integral $\int_0^\infty |q^{-1}\nabla_q V_{qu}|dq$ converges [3].

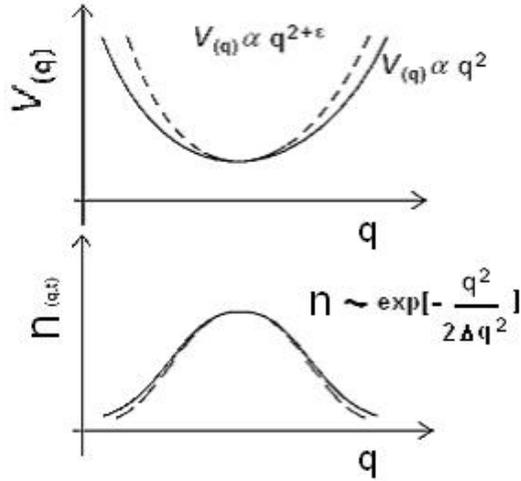
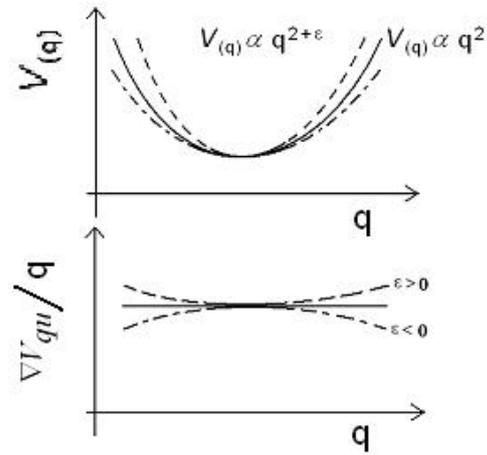

Fig.2 Particle density n for near-linear hamiltonian potential

Fig.3 The normalized quantum force $q^{-1}\nabla_q V_{qu}$ for near-linear hamiltonian potential

In this case, the mean weighted distance

$$\}_q = 2 \frac{\int_0^\infty |q^{-1}\frac{dV_{qu}}{dq}|dq}{\}_c^{-1}|\frac{dV_{qu}}{dq}|_{(q=\}_c)}}, \qquad (9)$$

evaluates the quantum potential range of interaction.
Faster the Hamiltonian potential grows, more localized is the PD and hence stronger is the quantum potential (fig.2 and fig.3). For the linear interaction, the Gaussian-type eigenstates lead to a quadratic quantum potential [6] and to a linear quantum force (i.e., $lim_{|q|\to\infty} |q^{-1}\nabla_q V_{qu}| \propto constant$) so that $\}_q$ diverges. A force weaker than the linear one as the Lennard-Jones type potential leads to $\}_q$ finite.

*2.2. Local stochastic dynamics*
When $\}_c \ll \Delta L$ and $\}_q \ll \Delta L$, where $\Delta L$ is the physical length of the system, the classical (local) stochastic dynamics can be achieved. In fact, given the condition $\}_q \ll \Delta L$ so that it holds $lim_{q\to\infty} -\nabla_q V_{qu} = 0$, $\dot{q}$ in (4) reads [3]

$$\dot{q} = \frac{p}{m} = \nabla_q \{ \lim_{\Delta L/\}_q \to 0} \frac{1}{m} \int_{t_0}^{t} dt ( \frac{p \cdot p}{2m} - V_{(q)} - V_{qu} ) \} = \frac{-\nabla_q V_{(q)}}{m} + \frac{\textcyr{u}p}{m} \cong \frac{p_{cl}}{m} \qquad (10)$$

where δp is a small fluctuation of momentum.

### 3. Discussion and conclusions

The large scale dynamics of (10), achieved for weaker-than-linear interacting particles in presence of the noise ɣ obey to the classical stochastic mechanics that, in a system of a huge number of particles, own a chaotic evolution. As widely accepted [1], such chaotic kinetics generate mixing in the phase space that brings the system to relax toward the thermodynamic equilibrium (or to a stationary state). Being irreversible, this evolution owns the so called macro-arrow of time. On the other hand, even if the quantum equation of motion (7) on a microscopic scale is indistinguishable by deterministic one, for $\Theta \neq 0$ it contains an infinitesimal presence of noise. If the standard quantum mechanics owns the time reversal symmetry, for $\Theta \neq 0$ the SHQA equation (1) or the SSE (7) do not (see fig.4).

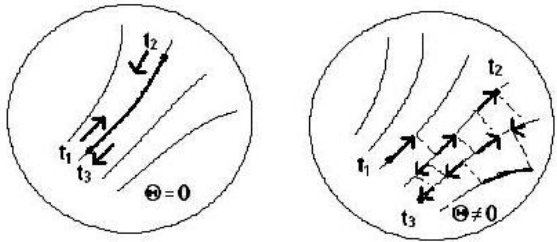

Fig.4  The forward and backward evolution along the SQHA phase space trajectories for an infinitesimal interval of time (on the left for $\Theta = 0$) and for $\Theta \neq 0$ (on the right).

For $\Theta \neq 0$ the time reversal breaking is not null and increases as $\Theta$ does. Since for the realization of the quantum mechanics we must be close to the deterministic limit $\Theta = 0$, the time symmetry breaking in (1,7) is expected to be small. Under the light of the SQHA approach, the logical separation between the concepts of micro-arrow and macro-arrow of time has no basic motivation: The residual and imperceptible presence of noise in the microscopic quantum dynamics leading to the micro-arrow of time is the premise for the breaking of the quantum coherence on macroscopic-scale physics and to the realization of the irreversible behavior owing the macro-arrow of time. Given the experimental evidences of time reversal asymmetry in elementary particle physics, the development of the relativistic hydrodynamic quantum analogy and its stochastic version can be the fundamental steps in obtaining, by a systematic procedure, the stochastic Dirac equation able to account for the time asymmetry in quantum relativistic physics.